\documentclass[12pt]{iopart}

\usepackage[final]{epsfig}
\usepackage{amssymb}
\usepackage{graphics}
\usepackage{color}

\newcommand{\BE}{\begin{equation}}
\newcommand{\EE}{\end{equation}}

\begin{document}
\title[The distribution of landed property]{The distribution of landed property}

\author{Pavel Exner${}^{1,2}$, Petr \v{S}eba${}^{2,3,4}$ and Daniel
Va\v{s}ata${}^{2,5}$}

\address{$^1$ Nuclear Physics Institute, Academy of Sciences of the
Czech Republic, CZ-25068 \v{R}e\v{z} near Prague, Czech Republic}
\address{$^2$ Doppler Institute for Mathematical Physics and Applied
Mathematics, B\v{r}ehov\'{a} 7, CZ-11519 Prague, Czech Republic}
\address{$^3$ University of Hradec Kr\'alov\'e, V\'{\i}ta Nejedl\'eho 573
CZ-50002 Hradec Kr\'alov\'{e}, Czech Republic}
\address{$^4$ Institute of Physics, Czech Academy of Sciences,
Cukrovarnick\'{a}~10, CZ-18000 Prague 8, Czech Republic}
\address{$^5$ Department of Physics, Faculty of Nuclear Sciences and
Physical Engineering, Czech Technical University, B\v{r}ehov\'{a}
7, CZ-11519 Prague, Czech Republic} \ead{exner@ujf.cas.cz,
seba@fzu.cz, daniel.vasata@gmail.com}

\begin{abstract}
The distribution of property is established through various
mechanisms. In this paper we study the acreage distribution of
land plots owned by natural persons in the Zl\'{\i}n Region of the
Czech Republic. We show that the data are explained in terms of a
simple model in which the inheritance and market behavior are
combined.
\end{abstract}

\pacs{05.65.+b, 89.65.Gh} 

\maketitle

\section{Introduction}

Quantitative study of human behavior requires reliable data and
tools to analyze them. The latter improved dramatically in recent
years when entries in various databases became available in a
digital form. From the sociological point this opens way to new
highly interesting empirical findings and theoretical endeavors to
understand them.

One of the first publicly available records were the telephone
books which gave an opportunity to study the distribution of
surnames. Interpreting the results in a model setting, the recent
records found in the telephone book are understood as a result of
a long process of intermarriages and surnames inheritability,
hence the surname statistics refers to social and population
processes in the past \cite{DMZ}. In particular, it is known that
the numbers of individuals sharing the same surname follows a
scaling rule known as Zipf law. Let us arrange the surnames in the
descending order with respect to the number of entries found in
the telephone book, and let $N(1)$ denote the number of records
found for the most frequent surname, $N(2)$ the number os records
for the second widespread surname, etc.; then the Zipf law says
that $\log(N(k))\approx -c \log(k)$ with some constant $c$.

The mechanism of inheritance from a single parent appears not only
in the social context: similarly to surnames it applies to
nonrecombining alleles in the genom, hence it is not surprising
that there is a close link between the surname distribution and
the human genome \cite{DMZ2}. Furthermore, such a situation is not
restricted to surnames but applies also to groups of people
sharing a common native language, to the species diversity in
ecological systems and so on -- cf. \cite{sole}. It demonstrates
that cultural traditions are transmitted from the ancestors to the
descendants through a process analogous to the genetic heredity
and display a close parallelism between the cultural and
biological evolution -- recall that in the evolutionary psychology
these mechanisms are studied under the name ``memetics''
\cite{selfish, meme}.

In the field of economics the power law distribution is
traditionally named after the classical work \cite{pareto} of
Vilfredo Pareto; it is used to describe phenomena such as the
statistics of personal income, or the allocation of wealth in a
steady society \cite{huang, patri}, however, also fluctuations of
the stock prices or the land estate display such a behavior
\cite{sor, kai}.

Our aim in this paper is to combine these two aspects into a
single model capable of describing situations in which both the
ancestor--descendant dynamics and the market behavior occur.
Specifically, we will analyze the distribution of the sizes of
land plots owned by individuals. It is clear that -- in contrast
to a corporative ownership -- a natural person can acquire the
land either on the real estate market or to inherit it. On one
hand the land is transmitted from an ancestor his/her descendant
similarly like surnames, on the other hand it is a subject to
changes by acquiring/selling the land on the real estate market.
We are going to demonstrate that such a complex human behavior can
be reasonably well described by a simple mathematical model.

The empirical basis for our investigation are data obtained from
the \emph{Czech Office for Surveying, Mapping and Cadastre}
describing the present status of the legal relations to real
estate property in the Czech republic. This includes, in
particular, the information about the sizes, types, geographical
location, and owners of the individual land plots; here we focus
on the statistical distribution of the plot sizes owned by
individuals.

The idea to use cartographic data is not new. For instance, the
information on the spatial structure of the urban networks has
been used in \cite{lae} to analyze the distribution of vehicular
flows. There are also direct attempts to model future developments
of the land use -- in particular in the urban regions -- such an
information is of value especially for the real estate developers.
These models, however, are not simple -- see, e.g., \cite{white}
for a cellular automaton approach. Our intention in the present
work is more simple: we are going to describe the actual division
of the surface of (a part of) the Czech Republic territory among
individuals as a steady state resulting from a long-term process
of real estate inheriting and trading. Needless to say, we do not
care about the individual plot sizes, only about their statistical
properties.

Let us describe briefly the contents of the paper. In the next
section we present and discuss a simple land trading/inheritance
model. The resulting statistical properties of the plot sizes will
be then compared with the true cadastral data for the Zl\'{\i}n
region in Section 3.

\section{A simple real estate trading/inheritance model.}

To motivate the model, note first that the Czech cadastral
records, which we work with, do not contain the land plot history,
and consequently, the past real estate transfers cannot be
directly extracted from them. There are indirect clues that
contain information on carried transfers: any conveyance of real
estate for consideration is subject to the real-estate-transfer
tax. On the other hand, the inherited or donated realties are
liable to the estate and capital transfer tax. Unfortunately, this
tax is levied on all assets which makes the information about
inherited realties obscured. Nonetheless, the taxing information
represents the process of realty conveyance in the metric of the
current prices. Recall that the price metric was used, for
instance , to analyze the mechanism leading to the crash of the
Japanese land market at the end of eighties -- see \cite{kai2}. It
does not say much, however, about the actual  acreage of the plots
and about its historical development. Another factor which makes
its use questionable is that the land price is subject to
unpredictable fluctuations; to quote an example from the same
study, the mean price of one square meter of land in Japan
increased in the years 1983 -1993 nearly sevenfold \cite{kai2}.
The dynamics of land prices exhibits a volatility which makes it
difficult to describe -- we refer to \cite{leu} and references
therein for more details.

With these facts in mind, we will focus on the statistical
properties of the acreage of the land plots (cadastral units).
They represent better objects to study because the plot sizes
change little in the course of time and do not yield to market
fluctuations. Most often the whole plot is conveyed and its size
does not change at all; changes of the plot size are rare and are
always related to a new surveying. The latter is typically a
complicated and costly procedure, which is one more reason why the
acreage is not vulnerable to speculations.

We understand the recent records contained in the land registry as a
description of a steady state resulting from a long series of land
inheritance and land trading. The total acreage is preserved, of
course, and can be only redistributed among the new owners. We are
going to describe the steady state by an agent-based approach
developed originally to model the wealth of closed economies -- see
\cite{lux} for a review. For the sake of simplicity we will use the
most elementary version designed initially to describe the social
stratification -- see \cite{patri} -- it is also known as the
``inequality process''.

As usual in  such situation we use discrete time proceeding in steps
typical for an ownership change; one can think of them as of
generations. Let $S_k(n)$ denote the acreage of the cadastral unit
$k$ at time $n$ and $S_{k+1}(n)$ be the the acreage of its
geometrical neighbor, by that we mean that $S_{k}(n)$
and$S_{k+1}(n)$ have a common border. Then the model cadastral
dynamics we propose proceeds to the next generation in the following
way,
\begin{eqnarray}
  S_{k}(n+1) &=& \lambda S_{k}(n) +a\left((1-\lambda)S_{k}(n)
  +(1-\mu)S_{k+1}(n)\right) \nonumber \\ [-.5em] \label{dynam} \\ [-.5em]
  S_{k+1}(n+1) &=& \mu  S_{k+1}(n) +(1-a)\left((1-\lambda)S_{k}(n)
  +(1-\mu)S_{k+1}(n)\right) \nonumber
\end{eqnarray}
where $\lambda, \mu \in [0,1]$ are independent random variables with
identical  distributions and $a$ is a Bernoulli variable taking
values $a=0,1$, each with probability $1/2$. The meaning of the
dynamical equations (\ref{dynam}) is straightforward. With
probability $1/2$ we have $a=1$ in which case the size of the plot
$k$ increases, $S_k(n+1)=S_k(n)+(1-\mu)S_{k+1}(n)$, while the size
$S_{k+1}$ of the unit $k+1$ is reduced, $S_{k+1}(n+1)=\mu
S_{k+1}(n)$. In other words, the cadastral unit $k$ of original
acreage $S_k$ has incorporated a part of the neighboring land plot
$k+1$. As a result, we get (after a surveying and introducing into
the land register) two new cadastral units of acreages $S_k(n+1)$
and $S_{k+1}(n+1)$. With the same probability we have $a=0$ in which
case the plot $k$ shrinks, $S_{k}(n+1)=\lambda S_{k}(n)$, with a
part of it being incorporated into $k+1$ giving
$S_{k+1}(n+1)=S_{k+1}(n)+(1-\lambda)S_{k}(n)$.

We are interested  in the probability distribution of the steady
state to which $S_k(n)$ converges as $n\to\infty$. A general
mathematical result ensures that the limit of such a process exist
and is unique -- see \cite{luc}. It depends, of course, on the
statistical properties of the variables $\lambda$ and $\mu$ and can
be numerically evaluated by iterating the mapping (\ref{dynam}) --
cf. \cite{knap}. Moreover, in analogy with \cite{garden} one can
argue that the convergence is fast coming close to the equilibrium
in just a few generations, hence one can expect the model will be
applicable to processes of land plot redistribution provided they
run undisturbed for at least a century.

The variable $\lambda$ describes the  land conveyance and contains
in such a way the information about the generation dynamics (we
refrain from mentioning the variable $\mu$ all the time since it has
identical statistical properties). The mechanism is simple: assume
that the owner of a cadastral unit has three offsprings. When he or
she dies, the children become after the appropriate inheritance
procedure co-owners of the cadastral unit with the one-third share
each; the fact is recorded into the land register. We disregard the
possibility that the heritage involves a more complex settlement
leading to other proportions assuming that such effects will
statistically cancel. Assume now that one of the descendants decides
to individualize his or her share, i.e. to discontinue the
co-ownership and to register the part as an individual belonging, or
he/she decides to sell it. After the appropriate surveying the
original size of the plot changes by the factor $1/3$, i.e. we have
$\lambda=1/3$ in this case. It has to mentioned here that matrimony
does not establish a joint property in the above judicial sense. A
married couple acquires a land estate into a tenancy in its
entirety. This is a different legal category and it is registered
differently; we will handle it as a full ownership of the plot. on
the other hand, a marriage separation  leads to a co-ownership.

An instantaneous snapshot of all these processes is recorded in
the cadastral database. On particular, we can derive from it the
probability $p_{n,m}$ that a fraction $n/m$ with $m>n$ of the
cadastral unit is in co-ownership. To take then into account the
genealogical part od the land conveyance we use this information
to define the appropriate distribution of the statistical variable
$\lambda$: we suppose that it takes the value $\lambda=n/m$ with
probability $p_{n,m}$ whenever $m\ge 2$. We can leave out the case
$m=1$ when the land plot is inherited as a whole, $S_k(n+1)
=S_k(n)$, noting that the identity of the owner is not important.

The inheritance does not tell us the whole  story, of course, since
a part of the real estate is traded on the market. This concerns, in
contrast to the above, the plots with $n=m=1$ having a sole
proprietorship with $\lambda=1$.  Such plots -- being fully owned by
a single person -- are subject to free trading. For simplicity we
will assume that the  trading mechanism is completely random in full
analogy with the closed economy models \cite{lux}). In other words,
we suppose that on the subset of plots with $n=m=1$ the quantity
$\lambda$ is a uniformly distributed random variable.

To summarize this part, we have formulated the  model of exchange of
cadastral units governed by the dynamical map (\ref{dynam}). For
units that are in a co-ownership of different persons the variable
$\lambda$ equals to the corresponding share and enters the equations
with the probability matching the relative appearance of the given
share among all the cadastral records. On the other hand, for
cadastral units have a sole proprietorship the variable $\lambda$ is
random and uniformly distributed. In such a way if a piece of real
estate is co-owned only the appropriater share can be traded, while
those with a single proprietorship can be on the other hand traded
without limit in a fully random way.

The corresponding dynamics can be now investigated  numerically.
One finds that the process converges fast, as we argued above, and
the resulting equilibrium distribution is not sensitive to the
initial state of the cadastre. In the next section that will
compare the steady result of this linear model with the data and
show that it leads to an amazing agreement with the acreage
distribution extracted from the Czech Land Registry.

\section{The results}

To begin with we note that if the random part of the dynamics
(\ref{dynam}) is omitted the land possession becomes fractalized.
As a mathematical result about iterative maps with Bernoulli
variables it was demonstrated not so long ago -- see, for
instance, \cite{BR01}. The essence of the effect, however, was a
fact well known to our ancestors and it was behind various
juridical acts attempting to prevent a land structure
disintegration. One can mention the so-called birthright edict --
a part of the legal reform of the Habsburg Emperor Joseph II dated
1787 -- which did not allow to split the land below 40
``measures'', which is roughly 7.6 ha or 19 acres. This law was
valid until the year 1869 when the rural land market --
representing the random part of the dynamics (\ref{dynam}) --
became vivid enough and the edict became obsolete. The imperial
act allowing the free divisibility and exchangeability of land (in
Moravia, from which our data are taken, it was valid already one
year earlier, in 1868) caused an unprecedented increase of the
land exchange. For example, according to \cite{Urban} in average
$10\%$ of estates changed the proprietary relations yearly during
the period 1869--87.

What we find in the present cadastral data is thus a result of a
process that started in 1868, some five generations ago. At the
beginning it was marked by an boosted land exchange and the
dynamics became steady only about a generation later. While these
fact are documented, one may speculate about other factors having
impact of the dynamics, in particular, changes in the population
structure. It is obvious that that the natality and the number of
offsprings living up to the adulthood were considerably higher at
the beginning of the process, and as a result, the distribution of
the land co-ownership was different from the present data.
Unfortunately, the digitalized Czech land registry does not
contain such a history record -- all we see is a snapshot of the
recent situation. In the light of these comments indicating how
simplified our model dynamics (\ref{dynam})) is, it is surprising
and worth of attention how well do its result agree with the
observed acreage probability distribution.

One more comment is needed before the results can be presented. We
are dealing with all the plots contained in digitalized cadastral
map \emph{with the exception of the build up areas}. The reason for
this methodical choice is simple: build up areas are places where
building stand. In the cadastral map, however, we see only the
ground plane, i.e. the projection of this building which does not
tell us what the latter is like: it can be a simple one floor house
as well as a high-rise construction. An information on the buildings
is not a part of topographical data collection being contained in a
separate registry, and some parts of the building (flats) can be
traded leaving the ground plan unchanged. This is why we exclude the
build up areas from the considerations and will discuss them in a
separate paper.

For the comparison we have used the cadastral records of the
Zl\'{\i}n Region, a territorial unit of the Czech republic of the
area 3964 $\mathrm{km}^2$ and population of 590 thousand. We
selected all the plots owned by individuals, excluding those owned
by companies. It is clear that the plots owned in a co-ownership,
or in the so-called undivided co-ownership of spouses are recorded
repeatedly times in the data, hence the duplications were removed
to get a meaningful size distribution. As indicate above all the
build up areas were deleted from the sample. This left us with the
collection of 1200121 plots which we could use to work out the
size statistics and to evaluate the acreage probability density.

\begin{figure}
\begin{center}
  \includegraphics[height=9cm,width=15cm]{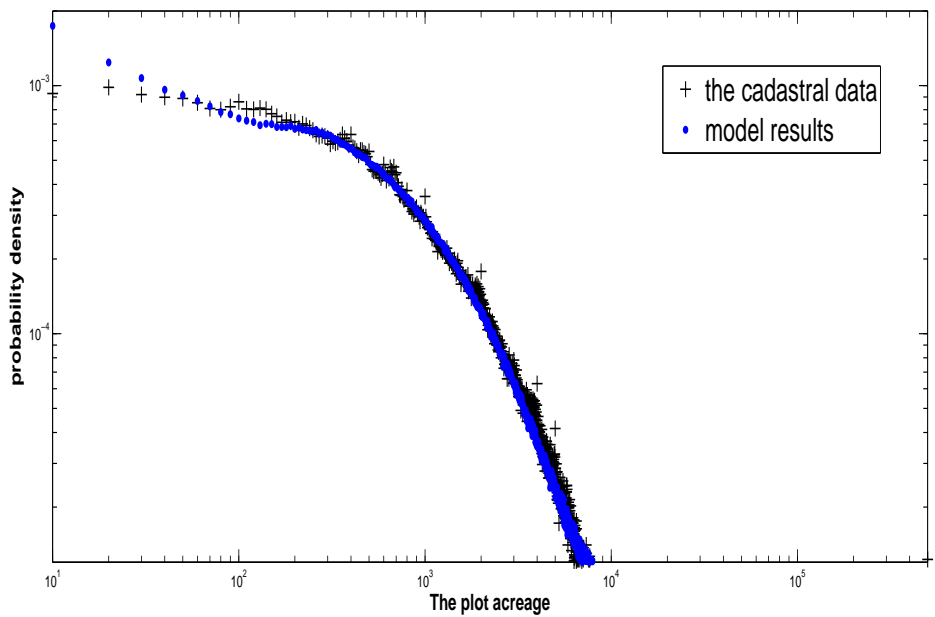}\\
\end{center}
  \caption{
The probability density of the a given plot acreage found in the
cadastral records (black crosses) is compared with the
distribution obtained by iterating the map (\ref{dynam}) (blue
points).
  }
  \label{plot1}
\end{figure}

The same cadastral data were used to evaluate the probabilities
$p_{n,m}$ of the particular shares $\lambda=n/m$ describing the plot
co-ownership as described in the previous section. The map
(\ref{dynam}) was then iterated starting from the uniform
distribution, $S_{k}(1)=1$ for all $k$. The result was subsequently
rescaled to the mean which equals the mean plot size found in the
cadastre and the two probability densities were compared; the result
is plotted on the Figure~\ref{plot1}.

\begin{figure}
\begin{center}
  \includegraphics[height=9cm,width=15cm]{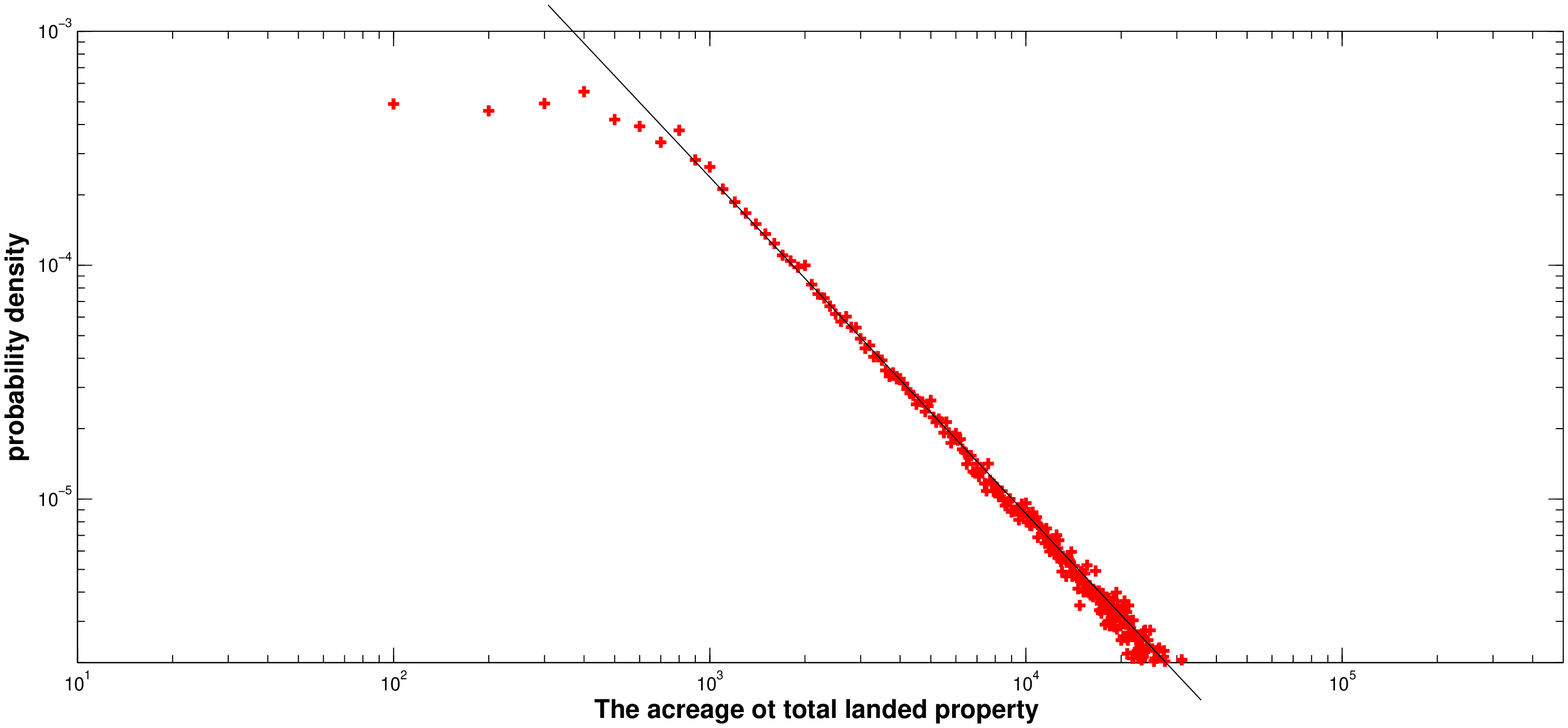}\\
\end{center}
  \caption{
The probability density describing the acreage of the total landed
property owned by individuals in the Zl\'{\i}n Region. The Pareto
behavior is  clearly visible
  }
  \label{plot2}
\end{figure}

We see that the two match perfectly; it is worth to stress that
the model \emph{contains no free parameters}. On the other hand,
it is obvious that the data do \emph{not} display the Pareto
behavior. This fact is easily explained; it is enough to realize
that the data represent the acreage probability of the plots and
not the the probability of the individual belongings. The point is
that one person can own more plots and their acreage (counting of
course with the related co-ownership fraction) has to be summed.
If we do this the Pareto behavior is restored as can be seen from
the appropriate distribution plot on the Figure~\ref{plot2}.

\subsection*{Acknowledgments}

The research was supported by the Czech Ministry of Education, Youth
and Sports within the project LC06002 and the Grant Agency of the
Czech Republic No. 202/08/H072. We are indebted to Helena
\v{S}andov\'{a} and Petr Sou\v{c}ek from the \emph{Czech Office for
Surveying, Mapping and Cadastre} for the help with acquiring the
data.


\vspace{3em}

\begin{thebibliography}{100}

\bibitem{DMZ}
Derrida B, Manrubia S C and Zanette D H 1999 \PRL \textbf{82}
1987-1990


\bibitem{DMZ2}
Manrubia S C and Zanette D H 2002 \emph{J. Theor. Biology}
\textbf{216} 461-477


\bibitem{sole}
Sole R V and  Alonso D 1998 \emph{Adv. Complex Systems} \textbf{1}
 203-220


\bibitem{selfish}
Dawkins R 1976 \emph{The Selfish Gene} (Oxford: Oxford University
Press) chap.~12


\bibitem{meme}
Blackmore S 1999 \emph{The Meme Machine} (Oxford: Oxford
University Press)


\bibitem{pareto}
Pareto V 1897 \emph{Course d'economie politique} (Lausanne: Rouge)


\bibitem{huang}
Huang Z F and Solomon S 2002 \emph{Physica} A \textbf{306} 412-422


\bibitem{patri}
Patriarca M,  Chakraborti A, Heinsalu E and Germano G 2007
\emph{Eur. Phys. J.} B \textbf{57} 219-224


\bibitem{sor}
Sornette D 2002 \emph{Physica} A \textbf{309} 403-418


\bibitem{kai}
Kaizoji T 2005 \emph{Physica} A \textbf{347} 575-582


\bibitem{lae}
Laemmer S, Gehlsen B and Helbing D 2006 \emph{Physica} A
\textbf{363} 89-95


\bibitem{white}
White R and Engelen G 2000 \emph{Computers, Environment and Urban
Systems} \textbf{24} 383-400


\bibitem{kai2}
Kaizoji T and Kaizoji M \emph{Physica} A \textbf{344} 138-141


\bibitem{leu}
Leung C K Y and Chen N K 2006 \emph{J. Real Estate Research}
\textbf{28} 293-320


\bibitem{lux}
Samanidou E, Zschischang E, Stauffer D and Lux T 2007 \emph{Rep.
Prog. Phys.} \textbf{70} 409-450


\bibitem{luc}
Devroye L and Neininger R 2002 \emph{Adv. Appl. Probability}
\textbf{34} 441-468


\bibitem{knap}
Knape M and Neininger R 2008 \emph{Methodol. Comput. Appl.
Probability} \textbf{10} 507-529


\bibitem{garden}
Exner P. and \v{S}eba P 2008 \JPA \textbf{41} 045004


\bibitem{BR01}
Baron M and Rukhin A L 2001 \emph{Stat. and Probab. Lett.}
\textbf{55}  29-38


\bibitem{Urban}
Urban O 1978 \emph{Capitalism and Czech Society} (in Czech;
Prague: Svoboda)




\end{thebibliography}
\end{document}